# Dislocation healing during hydrogen absorption and desorption in palladium

T.A. Webb[a,b], C.J. Webb[a], E.MacA. Gray[a,*]

[a]Queensland Micro- and Nanotechnology Centre, Griffith University, Brisbane 4111, Australia.

[b]Rocket Lab New Zealand, Mount Wellington, Auckland 1060, New Zealand.

**Abstract**

An *in-situ* neutron diffraction investigation of the annealing and healing of dislocations in the bulk-Pd–$D_2$ system was carried out. Lattice misfit between the α and β hydride phases produces dislocations during the phase transition in either direction, relieving elastic strain, which is reflected in reduced pressure hysteresis compared to the spinodal hysteresis. The effects on the dislocation density of annealing the metal under vacuum, of annealing in the β hydride phase, and of the phase transformation itself were investigated by measuring diffraction peak breadths during annealing and hydrogen cycling. During annealing under vacuum the dislocations were removed at a lower temperature than was previously reported, but annealing in the β phase gave nearly the same result. However, when cycling hydrogen in and out of the sample, the dislocation density decreased much faster with increasing temperature compared to annealing. In other words the process of phase transformation allows for healing of dislocations at lower temperatures than would be required to anneal them purely by heating. This healing effect was observed during both absorption and desorption. This result illuminates the mechanism by which misfit dislocations can be healed at the same rate that they are created in a sample undergoing absorption–desorption cycling, as proposed in theoretical models of the origin of pressure hysteresis.

# 1. Introduction

Dislocations are created in many metal hydrides during hydrogen absorption and desorption as a result of lattice mismatch between the $\alpha$ (dilute solid solution) and $\beta$ (concentrated hydride) phases. In bulk Pd, dislocation generation ceases at the critical temperature ($T_{crit}$ = 563 ± 1 K for Pd–$H_2$ and $T_{crit}$ = 556 ± 1 K for Pd–$D_2$ [1]), above which there is no phase transition. The role that the α–β phase transition plays in the hydrogen absorption properties of metal hydrides has been the topic of much discussion, with many questions still to be conclusively answered. This study investigates the creation and annihilation of dislocations during hydrogen cycling and provides insight into their relationship with pressure hysteresis in larger-than-nanoscopic particles. Pressure hysteresis, observed as a higher hydrogen gas pressure required for absorption ($p_{abs}$) compared to desorption ($p_{des}$), represents an energy loss proportional to ln ($p_{abs}/p_{des}$) in the storage process and may negatively affect the practical operation of metal-hydride storage units if $p_{abs}$ becomes too high compared to $p_{des}$ [2]. Models of hysteresis based on energy loss through dislocation generation [3] [4] [5] or, alternatively, on the thermodynamics of phase transformations [6] [7] appeared as long ago as 1937.

The earliest diffraction measurements with x-rays revealed the existence of the α and β phases, and that these phases retain the basic face-centred cubic structure of the metal [8]. The Pd unit cell parameter increases by approximately 3% during the α-to-β phase transformation at room temperature, causing the generation of misfit dislocations [9]. Dislocations require energy to create and this energy is dissipated as heat (phonons) during both the formation and the annihilation of the dislocations [3]. Since this heat is lost and cannot be recovered, pressure hysteresis must be present whenever dislocations are created [3] [5]. It has been reported that the misfit strain between the α and β phases in Pd causes dislocations to form during both absorption and desorption [10]. Jamieson *et al.* [11] investigated α phase growth and dislocation production in $PdH_x$ by loading a Pd sheet with hydrogen above the critical point (to avoid dislocation production) then cooling to room temperature and observing the sample during desorption of hydrogen in an electron microscope. They observed significant dislocation production as the metal plastically deformed at the advancing phase boundary when the α phase grew in β phase.

Ubbelohde [4] argued that strain arising during the formation of the β phase caused the absorption pressure to be higher than in the absence of strain, and that on desorption any strained areas would desorb first, leaving the desorption plateau as the true equilibrium. Wicke and Blaurock [1] found that if the plateau pressure was plotted against temperature, the desorption pressure followed the same trend below the critical point as the plateau pressure above the critical point, but that the absorption pressure did not follow this trend. This provided evidence to suggest that the desorption pressure was the "true" equilibrium pressure and the absorption pressure was somehow modified. However, Buckley *et al.* [2] carried out a more thorough investigation on LaNi$_5$ and found that the ratio of the plateau pressures for the first desorption and the second desorption was the same as the ratio of plateau pressure for the second absorption and the third absorption. The first absorption was not compared because of the activation effects during the first traversal of the two-phase region. This result suggests that neither absorption nor desorption should be regarded as closer to a "true" equilibrium pressure.

Lacher [6] developed a basic thermodynamic model for hysteresis in a two- phase hydride system. This theory proposed that absorption proceeded *via* a super-saturated metastable α phase and that desorption proceeded *via* an under-saturated metastable β phase. Lacher proposed that the free energy depends on the interfacial area between the α and β phases and that no other modification to the Phase Rule is necessary. This theory predicted that the "true" equilibrium pressure was half way between the absorption and desorption pressure. Schwarz and Khachaturyan [7] developed a model of hysteresis based around the thermodynamics of a crystal with coherent phase boundaries between the α and β phases. They modified the Phase Rule to include a term for long-range strain, which would be caused by the phase boundaries. They argued that any long-range strain that affects the free energy would give rise to hysteresis. However, in most real metal hydride systems, and particularly in bulk-PdH$_x$, dislocations are generated during the phase transformation, which implies that the phase boundaries are incoherent [11], at least in the two-phase region.

Griessen *et al.* [12] collated a large body of experimental data on Pd nanoparticles and extended the work of Schwarz and Khachaturyan [7] and Lacher [6] to apply a consistent coherent strain model to the data. They found that the data were fitted remarkably well by the model and concluded that hysteresis in Pd nanoparticles can be explained very well assuming between 45% and 100% spinodal hysteresis. They concluded that absorption proceeds by a path very close to the full spinodal pressure, while the desorption pressure is somewhere between the lower spinodal pressure and the Maxwell pressure. This model does not apply to bulk Pd because dislocations are formed before the spinodal concentrations are reached, relieving the stress and causing the phase transformation to be incoherent [12]. Therefore the presence of dislocations must allow easier accommodation of the α–β phase transition. This may be because the dislocations glide at the advancing phase boundary, or simply because the imperfect crystal accommodates strains more easily. What is not clear is whether the dislocations which were formed in the first cycle remain during subsequent cycling, implying that hysteresis in subsequent cycling is due to coherent strain with lower spinodal hysteresis, or whether dislocations are continuously created and removed in each cycle. What is clear is that the density of dislocations does not continually increase but reaches a maximum after perhaps just one absorption in Pd or a number of initial cycles in a brittle host such as $LaNi_5$ [3]. Given that hysteresis remains even after many cycles, it is not clear whether dislocations are still being created, or whether the hysteresis is purely due to coherent strain. If there is plastic deformation in every cycle, it is possible that the dislocation density saturates, as is the case in a metal which is continually cold worked. This is plausible since the dislocation density in Pd which has been hydrogen cycled at room temperature is similar to the dislocation density in cold-worked Pd. However, the dislocation density in hydrogen-cycled Pd has in some cases been reported to be higher than in cold-worked metals [13] and hydrogenation leads to a very different dislocation substructure [14]. There are therefore clear differences between the creation of dislocations by cold working and the creation of dislocations by hydrogen cycling. Alternatively, it could be that the dislocations are removed at the same rate at which they are created, as proposed by McKinnon [3]. Although the dislocations in hydrogen-cycled Pd were reported to be removed only after annealing above 400 °C [10], the presence of hydrogen and the process of the phase transformation could aid in the removal of dislocations, through enhanced dislocation mobility [15].

Ulvestad and Yau [16] observed dislocation healing during desorption by $PdH_x$ nanoparticles with size about 400 nm, which exceeds the reported critical size for dislocation generation of 300 nm [17]. The healing phenomenon was only observed during desorption. Dislocations were introduced during absorption and removed by the subsequent desorption.

Further interest in this problem was generated recently by Rahm, *et al.* [18], who carried out first-principles calculations of phase boundaries, pressure hysteresis and coherent interface energies for PdH$_x$ nanoparticles. In the present context, the key prediction was that between about 400 K and the calculated critical point (540 K compared to the experimental value of 563 K [1]) the phase transition proceeds via a single phase, therefore requiring no generation of misfit dislocations, while exhibiting hysteresis nonetheless.

Annealing a metal by dislocation removal is a complicated process, even in materials that have not been hydrogen cycled, owing to pinning and tangling. In a typical material, dislocation motion is restricted by defects (vacancies, interstitial atoms, precipitates) which pin the dislocations, preventing them from gliding to the edge of the grain, or by tangling with other dislocations or dislocation loops. Therefore removal of dislocations is strongly dependent on the energies required to anneal the defects pinning the dislocations or jump the pinning barrier, resulting in a range of temperatures being required for dislocation removal [19]. The removal of dislocations in a metal hydride is further complicated by the presence of hydrogen and by the phase change which typically accompanies the absorption/desorption process.

In this investigation we set out to test the dependence of dislocation dynamics on the phase of the hydride/metal and the process of the phase transformation. This was done by observing diffraction peak broadening owing to dislocations in PdD$_x$. High-resolution neutron diffraction measurements were made *in situ* while heating the sample under vacuum and in the β hydride phase, and during absorption/desorption cycles at varying temperatures.

## 2. Materials and Methods

The sample used was Goodfellow Pd powder with 99.95% purity and 150 $\mu$m maximum particle size, which is macroscopic as far as external dimensions are concerned, although it is worth noting that Wu *et al.* [20] measured a "particle size", meaning coherently diffracting domain size, of 145 nm for the same sample material, well into the size range of interest in published studies of nanoscopic Pd hydrides.

The sample was measured in a 10 mm ID custom thin-walled stainless-steel neutron cell. Time-of-flight (TOF) neutron diffraction profiles were collected *in situ* on the High Resolution Powder Diffractometer (HRPD) at the ISIS spallation source, Rutherford Appleton Laboratory, UK. Neutrons are advantageous compared to x-rays for studying bulk Pd because the penetration depth is much greater; for the 9-keV x-rays employed in [16], the penetration depth is only 540 nm.

The diffractometer peak shape was characterised using $CeO_2$ and the time-of-flight was calibrated to *d*-spacing using NIST 640c Si. Deuterium (D, $^2H_1$) was used in place of $^1H_1$ because it has a much lower incoherent scattering cross section and a higher coherent scattering cross section than $^1H_1$. A Sieverts-type gas handling apparatus was used to apply $D_2$ pressure to the sample. Diffraction profiles were collected for 45 to 60 min to obtain data with adequate statistics for reliable Rietveld profile refinement. Profile analysis was carried out using Topas Academic [21]. The (200) peak was chosen for comparison because it has the largest broadening for the primary dislocation type observed in hydrogen-cycled Pd [20]. The (200) peak was fitted with a Gaussian function, rather than a more general Voigtian, because the Voigtian introduced an extra fitting parameter without fitting the data any better. The full-width half-maximum (FWHM) of the peak was used to compare the results of the various treatments of the sample. Data from the 90° (medium resolution) detector bank of HRPD were used because it had a much higher count rate than the high-resolution backscattering detector bank.

The temperature of the sample was monitored by a thermocouple attached to the outside of the sample cell. The lattice parameters of the deuterium-free initial Pd sample were used to calibrate the measured temperature against temperature determined from the lattice parameter in the volume of sample illuminated by the neutron beam. This accounted for the temperature difference between the thermocouple and the sample. First, a room temperature lattice parameter for Pd was accurately determined. The lattice parameter for Pd at 20 °C (293.15 K) was found to be a = 3.89027(3) Å. We regard this as more accurate than the value of 3.8902(6) Å published by Arblaster [22]. Our new value for the absolute lattice parameter at 20 °C was combined with the relative thermal expansion data collated by Arblaster [22]. The rationale for this approach is that calculating relative values from an internally consistent data set with systematic errors partially cancels these errors. Finally, we measured the Pd lattice parameter at higher temperatures and used the published thermal expansion data [22] to calculate the true temperature in the illuminated volume of the sample. The absolute accuracy of this temperature measurement was estimated to be ±5 °C.

The sample had been desorbed previously at room temperature following absorption of D at 350 °C, and so already had a high initial dislocation density. To begin the first experiment, a diffraction profile was recorded at room temperature. The sample was hydrogenated by applying approximately 50 bar of $D_2$ pressure. The temperature and pressure were allowed to stabilise and then a diffraction profile was recorded. The sample was then placed under vacuum to desorb and a diffraction profile was recorded. This pressure-swing cycling process was repeated at 156 °C and then in steps of approx. 21 °C up to 322 °C. Higher $D_2$ pressures of 65/75 bar were used at 260/281 °C respectively, and 85 bar was applied at 301 °C and 322 °C, to ensure that the sample was hydrogenated well into pure β phase. Figure 1 summarises the experiment.

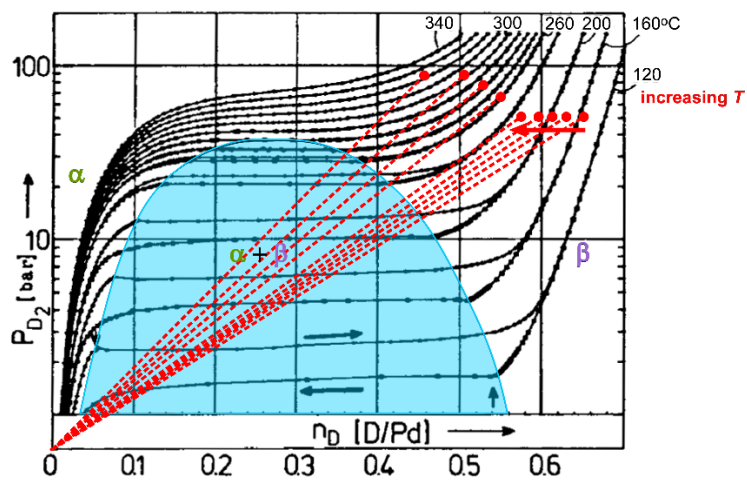

**Figure 1.** Phase diagram showing the measurement points for the first experiment, employing pressure-swing cycling at increasing temperatures (red arrow). The red dashed lines join the extreme points of each excursion in pressure and composition. Subject to temperature changes during rapid absorption and desorption, each excursion followed approximately the isotherm corresponding to the red point in pure β phase. In Figs 1–3, the isotherms are reproduced from Wicke and Blaurock [1].

At each temperature, the real locus of $(n_D, P_{D_2})$ is somewhat above the corresponding isotherm during absorption and somewhat below it during desorption, owing to the temperature excursion caused by the reaction enthalpy.

The second experiment aimed to establish the effect of a high concentration of D on the rate of dislocation removal. An *in-situ* annealing experiment was carried out under both vacuum and deuterium pressure. D$_2$ at 56 bar was applied to the sample at room temperature, creating dislocations in the process. While maintaining the deuterium pressure, the sample was annealed *in situ* by setting the temperature, waiting 1 hour and then collecting a diffraction profile. This was done at temperatures from 208 °C up to 364 °C in roughly 52 °C increments. Five consecutive scans were taken at 364 °C to confirm that the dislocation density had equilibrated. Figure 2 summarises the experiment.

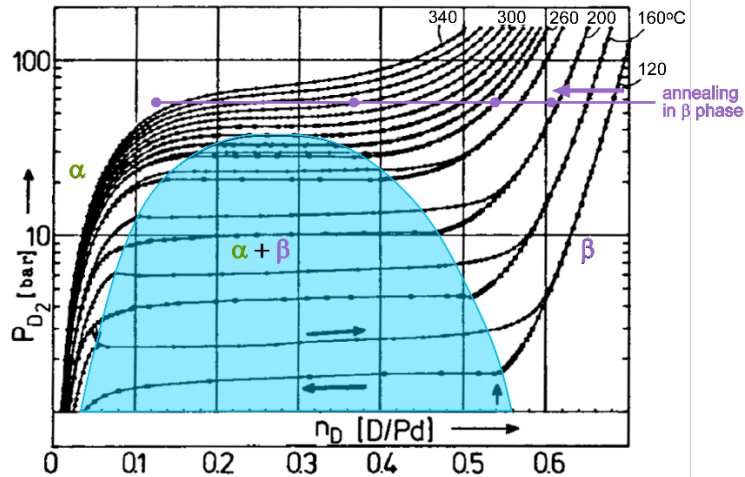

**Figure 2.** Phase diagram showing the measurement points for the annealing sequence beginning in pure β phase (second experiment). The arrow indicates the direction of the sequence from low to high temperature.

For the third experiment, the temperature was reduced to 104 °C and the sample was desorbed under vacuum, introducing new dislocations in the process. The annealing experiment was then repeated under vacuum at temperatures from 208 °C up to 520 °C in roughly 52 °C increments, taking four consecutive scans at 364 °C and seven at 520 °C to confirm equilibration.

Prior to the fourth experiment, the sample was removed from the cell and annealed in a vacuum furnace at approximately 700 °C for approximately three hours to remove defects. The sample was then reloaded into the sample cell and a diffraction profile was collected at room temperature. This experiment was analogous to the first experiment, except that a decreasing temperature sequence was followed during pressure-swing cycling. The temperature was raised to 322 °C (above the critical point at 283 °C [1]) and a diffraction profile was collected. The measurement cycle was: hydrogenate the sample by applying approximately 75 bar D$_2$ pressure; wait 5–10 min for temperature equilibration; collect a diffraction profile in pure β phase; evacuate the cell to desorb the sample; wait for equilibration and collect a diffraction profile. This was repeated at 291, 270, 249, 229 and 197 °C. Figure 3 summarises the experiment.

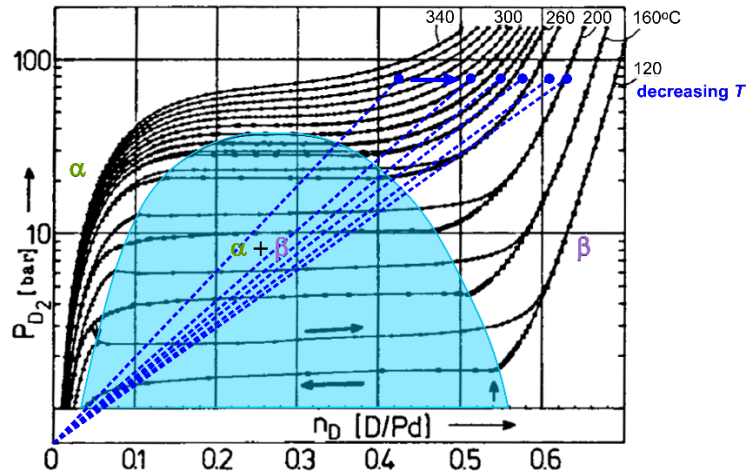

**Figure 3.** Phase diagram showing the measurement points for the fourth experiment, employing pressure-swing cycling at decreasing temperatures (blue arrow). The blue dashed lines join the extreme points of each excursion in pressure and composition. Each excursion followed approximately the isotherm corresponding to the blue point in pure β phase.

## 3. Results

The diffraction profiles obtained during pressure-swing cycling at increasing temperatures (first experiment) are shown in Fig. 4. The peak breadths for all dislocation removal experiments are shown in Fig. 5. When the sample was annealed under vacuum (third experiment), the peak breadth, and therefore the dislocation density, decreased with temperature as expected. However, at approximately 500 °C, the temperature used by McLennan *et al.* [23] to anneal Pd, the sample still contained a high dislocation density compared to the initial sample, even after several hours at this temperature. Annealing in the β phase (second experiment) showed no significant difference to annealing under vacuum, despite the presence of a high concentration of deuterium. This appears to be at odds with reports that the dislocation mobility significantly increases with the introduction of hydrogen to the sample [15], although the amount of dissolved hydrogen was much higher in the present case.

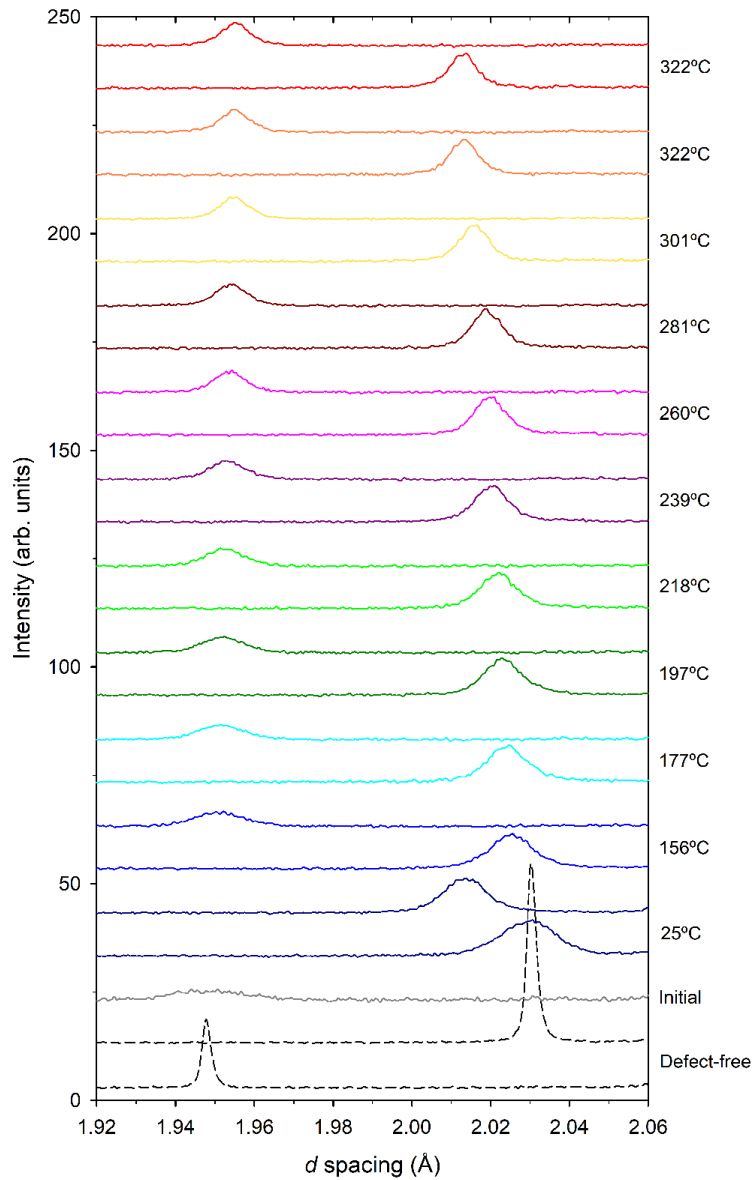

**Figure 4.** Diffraction profiles of PdD$_x$ showing the evolution of the (200) peak during pressure-swing cycling with increasing temperature (first experiment). The Initial profile was recorded after desorption at room temperature, as detailed in the text. At each temperature, the lower profile with higher $d_{200}$ was recorded after absorption and the upper profile with lower $d_{200}$ was recorded after desorption under vacuum.

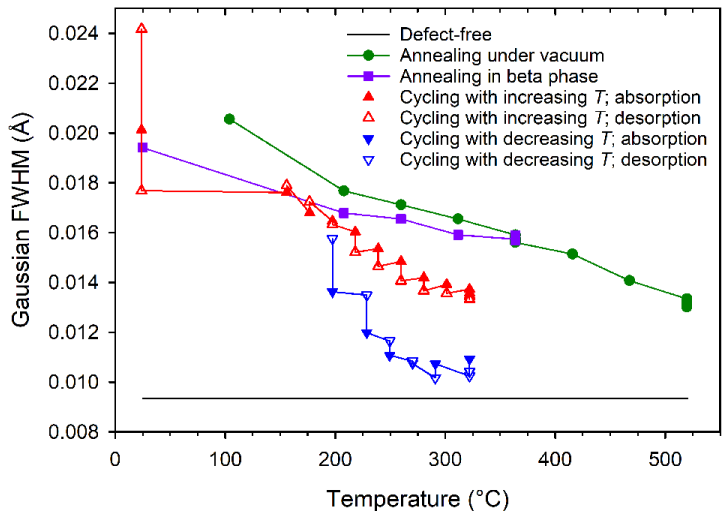

**Figure 5.** Temperature dependence of the breadth (FWHM as *d*-spacing) of the Gaussian fit to the (200) peak for annealing under vacuum, annealing in the β-phase, pressure-swing cycling with increasing temperature steps in the absorbed and desorbed states, pressure-swing cycling with decreasing temperature steps in the absorbed and desorbed states, and for reference, defect-free Pd.

Remarkably, pressure-swing cycling at increasing temperatures resulted in a much faster decrease in dislocation density compared to annealing the desorbed metal or β phase at the same temperature. This indicates that the process of phase transformation itself aids in the removal of dislocations, denoted healing by Ulvestad and Yau [16] to distinguish it from the usual static annealing process.

The first three points at room temperature are somewhat unexpected. The sample started with a very high dislocation density, but after absorption this decreased significantly. This means that the number of dislocations created during absorption was significantly lower than the number removed during this same step. One possible explanation for this is that the first traversal of the two-phase region in a defect-free sample creates a much higher dislocation density than subsequent traversals of the two-phase region. The third data point (the second point taken with the sample under vacuum) in fact corresponded to the sample being still in the absorbed state, because the kinetics of desorption at room temperature were too slow for the sample to desorb in the time allowed. Therefore it is hard to draw any conclusions about this data point, although it is surprising that it had a lower dislocation density than the previous point.

The healing effect during pressure-swing cycling at decreasing temperatures (fourth experiment) was even more marked: the dislocation density did not approach that of a sample cycled at low temperature until the temperature fell below approximately 200 °C, although there was a small but definite increase in dislocation density at 250 °C. It is interesting to note that in its initial state, the sample used for pressure-swing cycling at decreasing temperatures still contained a measurably higher dislocation density than the reference sample did. This shows that the *ex-situ* heating under vacuum for several hours at 700 °C did not completely anneal the sample.

## 4. Discussion

Given that the density of dislocations in metal hydrides does not increase with continued absorption/desorption cycles [3], as mentioned earlier, the two possible explanations are that the density of dislocations saturates or that the dislocations are removed at approximately the same rate they are created. The data in Fig. 5 show that in all cases the dislocation density started to decrease at much lower temperatures than the 400 °C previously reported for the removal of dislocations [10]. Furthermore, the dislocation density in the cycled sample reduced with increasing temperature even faster than during annealing in vacuum. The initiation of dislocation removal at a lower temperature is consistent with the authors' previous work on the kinetics of dislocation removal [24]. The pressure-swing cycling results provide evidence that the transformation between the α and β hydride phases in the $PdD_x$ system promoted dislocation removal (healing) at lower temperatures than would normally be required. The significant internal stress at the advancing α–β phase boundary may allow the structure to overcome potential barriers which would be too high for thermal fluctuation to overcome. As a result, the dislocations could be gliding at the interface between the phases as it advances through the crystal, allowing dislocations to glide to the edge of the grain and be removed. The advancing phase boundary could therefore allow motion and removal of dislocations, as well as creating new dislocations. Therefore, after a number of traversals of the two-phase region, the dislocation density would be expected to approach a stable value corresponding to that temperature. This is in fact exactly what is observed in pressure–composition isotherms of palladium hydride. This is a significant result because it provides insight into how the presence of dislocations accommodates the large internal strain and results in lower hysteresis in cycles subsequent to the activation cycle. This does not rule out the possibility that there may be a saturation effect of the dislocation density, but if this is the case it must be only one part of the picture. The gliding of dislocations at the front of the phase boundary would then be accommodating the strain of the phase transformation, resulting in the observed lower hysteresis for bulk Pd after the activation cycle, compared to Pd nanoparticles. The possibility that motion of the α–β interface might relieve dislocations was in fact raised by Timofeyev et al. [25] in 1980, in relation to their observation that Pd is more easily deformed following hydrogenation.

When pressure-swing cycling at decreasing temperatures, the dislocation density only increased after cycling at 250 °C. This could be because between 250 °C and 283 °C all dislocations created during the phase transformation were removed before a measurement was taken or because the increase in dislocation density created was too low to measure. The fact that the initial sample had dislocations present prior to the experiment would make the creation of low dislocation densities even harder to observe, so the latter possibility cannot be discounted. Whatever the details, it appears that in a range of temperatures below the critical point there is relatively little generation of dislocations, implying that the interface between the α and β phases is coherent or nearly so, as predicted for nanoparticles [7] [12].

Noting that the presented experimental results are for macroscopic rather than nanoscopic Pd particles, albeit with nanoscopic crystallite size, and that the crystallites are bounded by grain boundaries or similar structures instead of a free surface, the observed dislocation removal during hydrogen cycling is evidently the same healing phenomenon observed by Ulvestad and Yau [16] in nanocrystals, with the difference that it occurred during absorption as well as desorption. From the point of view of hysteresis and the existence of a mechanism to prevent dislocations building up, this symmetry makes sense.

In relation to the prediction by Rahm *et al.* [18], that in nanoparticles the phase transformation is not only coherent but occurs continuously in a single phase, the presented results support a coherent transformation but do not provide an answer as to the number of phases involved in the transformation, because diffraction profiles were recorded only at the start and end of the traversals of the two-phase envelope, not inside it.

Although the presence of a small amount of hydrogen is well known to promote dislocation motion [15] [26], and the presence of around 1% H has been shown to enhance the recovery of cold-worked Pd relative to vacuum annealing [26], annealing under vacuum and in the β phase gave very similar results (Fig. 5). There could be several reasons why dislocations were not observed to be removed any faster in the β phase for palladium hydride. The investigations of dislocation mobility were founded on softening of metals, i.e. the increase in dislocation motion with an applied stress. Although the reduction of the potential barrier for dislocation motion with an applied stress should also decrease the temperature required for annealing (since the barrier is lower compared to $kT$), this requires the assumption that annealing proceeds by the same mechanism as deformation caused by external stresses, which may not be the case. It was also reported that the softening of the metals in the presence of hydrogen only applied to the "high temperature" region, the absolute value of which is specific to the material [15]. It is therefore possible that the temperatures used in the current experiment may not have resulted in increased dislocation mobility. The quantity of hydrogen present may also have an effect on the results. It was reported that when the metal formed a hydride, no increase in dislocation mobility was observed [15]. It was suggested that this was because plastic deformation caused by the hydrogenation process created dislocations, obscuring any increase in dislocation mobility, however this was not conclusive. It is also possible that the small quantity of hydrogen which may have been trapped in the sample that was annealed under vacuum (since it was hydrogenated to create dislocations) may have had the same effect on dislocation mobility as in the sample annealed in the β hydride phase. This would therefore have resulted in the same annealing characteristics, both of which could possibly be enhanced compared to Pd which has not been exposed to hydrogen. Annealing in vacuum decreased the dislocation density somewhat above about 200 °C. This observation appears to be at odds with the conclusion by Sakaki *et al.* [10], based on positron annihilation results, that dislocation density in Pd starts to decrease at about 400 °C. While dislocations are not mobile by a glide mechanism at such a low temperature, vacancies are mobile and dislocation climb is a possible mechanism.

The results also reinforce the important point made by Buckley *et al.* [27] that the sample history has a significant effect on its microstructure and hysteresis, since hysteresis is affected by microstructure, specifically the existence of dislocations. This means that all experiments on hysteresis and microstructure must have carefully controlled sample preparation, since a relatively small change in temperature or a single traversal of the two-phase region can significantly change the dislocation density and the amount of pressure hysteresis. Comparisons of the plateau pressures of absorption and desorption are therefore somewhat difficult, since the sample history (of temperature and cycling) would have to be identical.

The presented experimental results allow a qualitative interpretation of some results in the literature for which no satisfactory explanation has been advanced.

Tripodi *et al.* [28] measured the resistivity of a 50-micron Pd wire subjected to 0.49 N tensile force ($\sigma \approx 260$ MPa). They observed that exposure to hydrogen greatly increased the strain rate of the sample, which deformed rapidly during the $\alpha - \beta$ phase transformation, leading to increased resistivity. The resistivity increase was only partially recovered during desorption. The results were ascribed to the "combined effect of mechanical stress and H dynamics".

In an analogous later study, Kawasaki *et al.* [29] measured the vertical deflection of a horizontal Pd plate with dimensions 70 l × 10 w × 1 t mm$^3$, anchored at one end, with a 124-g weight loading the free end. In the absence of hydrogen, the bending moment applied by the weight was too small to cause plastic deformation. At 150 °C deformation was observed during both absorption and desorption when the sample was exposed to 1 MPa hydrogen. When the hydrogen pressure was 0.2 MPa, however, a partial recovery (decrease of vertical deflection) was noted during desorption. The results were tentatively interpreted in terms of superplasticity.

These experiments can be interpreted based on the new understanding arising from the present work. Absorption or desorption of H by bulk Pd creates dislocations and greatly facilitates their motion while the phase transformation is proceeding. Easy dislocation motion greatly decreases the strength of the material, thus accumulating further plastic deformation from each passage through the two-phase region.

In the case of the Pd wire [28], the smaller increase of the resistivity during desorption was evidently a concatenation of a decrease as the H was desorbed and an increase owing to further plastic deformation. In the case of the Pd beam under 0.2 MPa hydrogen [29], the sample displacement decreased during desorption. This could only be effected by a force working against gravity, for which the only plausible candidate is tension in the top surface of the plate. It may be noted in Fig. 5 of Ref. [29] that the downward displacement of the weight halted when the pressure dropped below 0.17 MPa. This pressure is only marginally high enough to form the pure β phase at 150°C [30] in unstrained Pd. Under strain, the contracted lower portion of the bar would contain less hydrogen than the upper portion, causing the upper half to desorb first and shrink, possibly causing the upward "recovery" of the displacement. At 1 MPa charging pressure, the entire plate would have formed the β phase, so the "recovery" effect would not be expected.

## 5. Conclusions

The experiments reported here showed clearly the existence of a dislocation healing phenomenon in bulk polycrystalline $PdD_x$ at temperatures above about 225 °C. Contrary to previous experiments on nanoparticles, healing during passage through the two-phase envelope of the phase diagram occurred during both absorption and desorption. When absorbing/desorbing directly between fully desorbed Pd and the pure β phase (pressure-swing cycling), starting at a temperature well above the critical point, the dislocation density did not rise to the level normally found in desorbed Pd or the pure β phase until the temperature fell to about 200 °C. This suggests that the coherent phase transformation believed to occur in single Pd nanoparticles may actually occur in polycrystalline particles in a range of temperatures below the critical point.

In contrast, and despite the expectation of higher dislocation mobility in samples with hydrogen present, the β phase did not show any significant difference during static annealing compared with the sample annealed under vacuum. It is proposed that this could be owing to the pinning of dislocations by other defects which are not affected by deuterium, or the sensitivity to deuterium was high enough that the α phase was affected equally due to trapped deuterium.

To account for the dislocation healing phenomenon in bulk Pd, it is proposed that dislocations gliding at the front of the advancing phase boundary would result in their removal at the completion of absorption/desorption, when the dislocations would reach the edge of a grain. This mechanism would allow dislocations to be removed at the same rate they were being created during cycling, thus explaining how the dislocation density does not build up cycle by cycle. It also provides insight into how the presence of dislocations accommodates the large internal strain and results in lower hysteresis in cycles subsequent to the activation cycle, adding to the understanding of the relationship between structure and hysteresis in Pd, and more generally, in metal hydrides.

The experimental results also allow a qualitative interpretation of some formerly puzzling results in the literature that link passage through the two-phase region with decreased mechanical strength in PdH$_x$.

**Acknowledgments**

TAW acknowledges receipt of an Australian Postgraduate Research Award and thanks the Australian Institute of Nuclear Science and Engineering for financial support. The authors thank the Rutherford Appleton Laboratory, UK for the use of facilities and financial support, and K.S. Knight and D.S. Pyle for valuable assistance with the neutron diffraction measurements. EMG thanks N. Armanet for helpful discussions.